# MDPCT1 quench data and performance analysis

S. Stoynev, M. Baldini, E. Barzi, *Senior Member, IEEE,* G. Chlachidze, V.V. Kashikhin, S. Krave, I. Novitski, D. Turrioni, A. V. Zlobin

***Abstract*—MDPCT1 is a four-layer cos-theta $Nb_3Sn$ dipole demonstrator developed and tested at FNAL in the framework of the U.S. Magnet Development Program. The magnet reached record fields for accelerator magnets of 14.1 T at 4.5 K in the first test and 14.5 T at 1.9 K in the second test and then showed large degradation. While its inner coils performed exceptionally well with only two quenches up to 14.5 T and no evidence of degradation, the outer coils degraded over the course of testing. By adopting new measurement and analysis techniques at FNAL we are discussing in detail what happened. Both success and failure in our diagnostics are discussed. The evolution of techniques over the course of two tests (and three thermal cycles) shows the path to address challenges brought by the first four-layer magnet tested at FNAL. This paper presents the analysis of quench data along with diagnostic features and complementary measurements taken in support of the magnet performance analysis.**

***Index Terms*— Accelerator magnets, superconducting magnets, $Nb_3Sn$, current-voltage characteristics, acoustic measurements**

## I. Introduction

As part of the U.S. Magnet Development Program [1], the "15 Tesla" dipole demonstrator, a four-layer cos-theta magnet of ~ 1 m length called MDPCT1, was developed [2] and tested at FNAL. The work was conducted in collaboration with LBNL, and some coil parts were also provided by CERN as part of a long-standing collaboration. A similar strength dipole magnet was developed and tested in Europe [3]. MDPCT1 was initially tested to 14.1 T at 1.9 K and 4.5 K [4] after which its axial support for outer layers (outer coils) was modified and the magnet retested [5]. The reassembled magnet was called MDPCT1b. The maximum field at the magnet center reached 14.5 T at 1.9 K and next to the magnet lead-end it reached 14.6 T [5]. This is 97% of the magnet design field but its performance quickly degraded, especially after a thermal cycle (TC). MDPCT1 featured extra instrumentation and diagnostics compared to previous tests at FNAL. We temporarily lost some of it in the first test (voltage taps, acoustic readout from sensors on the end plates) and other was never recovered (strain gauges on the outer coils, pole areas). Bore temperature sensors and bore ("cold" and "warm") quench antennas (QA) were consistently reliable sources of data. Special low noise voltage measurements contributed to valuable insights on the conductor state. Here we report on investigations of magnet performance and quench behavior and briefly discuss prospects for further improvements of diagnostics.

This manuscript has been authored by Fermi Research Alliance, LLC under Contract No. DE-AC02-07CH11359 with the U.S. Department of Energy, Office of Science, Office of High Energy Physics; and supported by US-MDP (Corresponding author: Stoyan Stoynev).
All authors are with Fermilab National Accelerator Laboratory (e-mail: stoyan@fnal.gov).

## II. Quench Analysis

### A. Magnet Instrumentation and Diagnostics

The primary source of information for the quench analysis was voltage taps (VT) and their sequential segments. The VT schematics used in MDPCT1 is shown in Fig. 1. While most of the VT on the outer layers 3 and 4 became non-responsive early in the initial testing (first test), all but one were recovered for MDPCT1b. In addition, two types of quench antennas provided flux-change information: one inside the warm bore with seven axially placed elements [6] and another based on flexible Printed Circuit Board technology, flex-QA [7], with 2 x 6 sensors attached to the outer surface of the warm bore tube (opposite azimuthal locations), Fig 2. Two temperature sensors were also installed on the poles of the innermost layers, close to the magnet lead end. All those data were taken up to 1 s before quench detection at ~ 8 kHz acquisition rate.

Two acoustic sensors [8], fabricated at LBNL, were installed on each end plate, and data taken at up to 1 MHz acquisition rate during a whole current ramp. Most, though not all, ramps were covered in that regard. Pointedly, only MDPCT1b gave useful data due to earlier technical problems. Special low-noise voltage measurements were taken towards the end of the last thermal cycle, after all other "cold" tests with the magnet, with a multi-channel nano-voltmeter. Data were taken at various stable current levels with direct cable connection to the test stand instrumentation tree and used for V-I characterization of cable segments. Signal averaging over fast sampling extended to 1 s per measurement point. The device performance was verified by independent measurements of splice resistances with uncertainties below 0.1 nΩ level. As all but two quenches happened in the outer coils, where we lost all our pole strain gauges, we missed the opportunity to directly measure strains at those locations. The losses were associated to imperfect interfaces between edges of metallic coil parts and coil. Other strain gauges did not produce enough insights on magnet performance, and we do not discuss them.

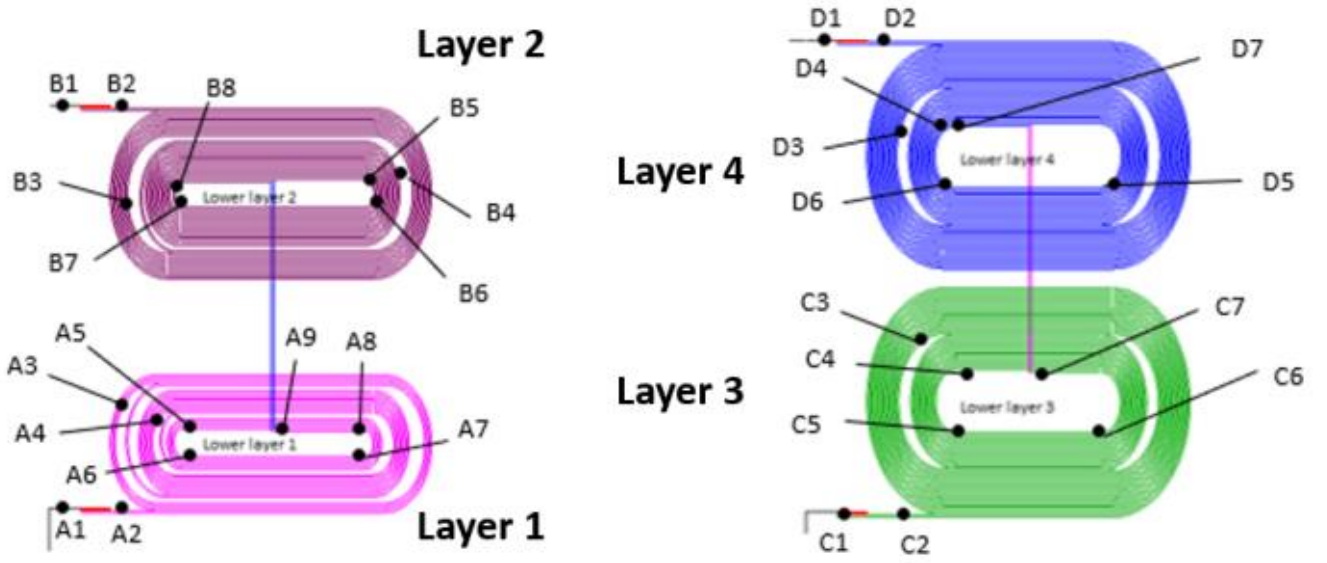

Fig. 1. Voltage Tap (VT) schematics in the "15 T" magnet. Inner coils 2 and 3 form layers 1 (A) and 2 (B), and outer coils 4 and 5 form layers 3 (C) and 4 (D). Readout is based on pairs of VTs (e.g., 5c6_c7 is a segment in coil 5 made of VTs c6 and c7). The lead end (LE) is on the left of the figures.

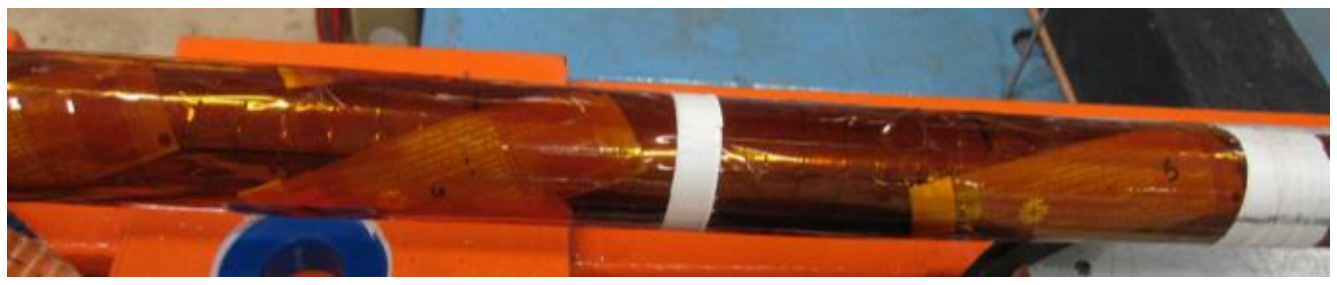

Fig. 2. Flex-QA elements installed on a warm bore tube.

## *B. Features in Magnet Training*

MDPCT1 training data were previously presented [5] but better understanding of features requires a more complete representation [9]. Short Sample Limits (SSL) for inner coils were extracted earlier and determined to be 10.8 kA / 12.2 kA at 4.5 K / 1.9 K. For outer coils those were 12.0 kA / 13.5 kA at 4.5 K / 1.9 K. Values for each coil pair agreed to within 0.1 kA. All training quenches were at 1.9 K and 20 A/s, special studies were performed in MDPCT1b only. From design perspective the inner coils are subjected to higher field and stresses [2],[4] and were expected to get most of the training quenches. Instead, out of more than 100 quenches only two were in the inner coils and one of them never quenched. Inner coils reached 82/90 % of their SSL at 1.9/4.5 K with little quenching which is extraordinary for $Nb_3Sn$ coils. Earlier $Nb_3Sn$ coils showed semi-independent training in magnets [9]. It is yet unclear if MDPCT1 being a four-layer magnet imposes a different behavior.

Both outer coils underperformed, with their performance affected by the initial imperfect end support and possibly design [10]. The training curve of MDPCT1b is shown in Fig. 3. The limiting outer coil reached ~74% of its SSL, both at 1.9 K and 4.5 K, and earlier both outer coils started training at 56-58% SSL. After reaching its record field in MDPCT1b, magnet quenching became erratic with signs of degradation. Temperature dependence studies clearly showed degradation with respect to the first magnet test [5] with nearly 10% lower current (field) at 4.5 K. This was followed by additional sudden 20% degradation after a thermal cycle [5] with reproducible and nearly identical currents as fractions of SSL at 1.9 K and 4.5 K. Virtually all features observed pertained to one of the outer coils – coil 5 (Fig. 3).

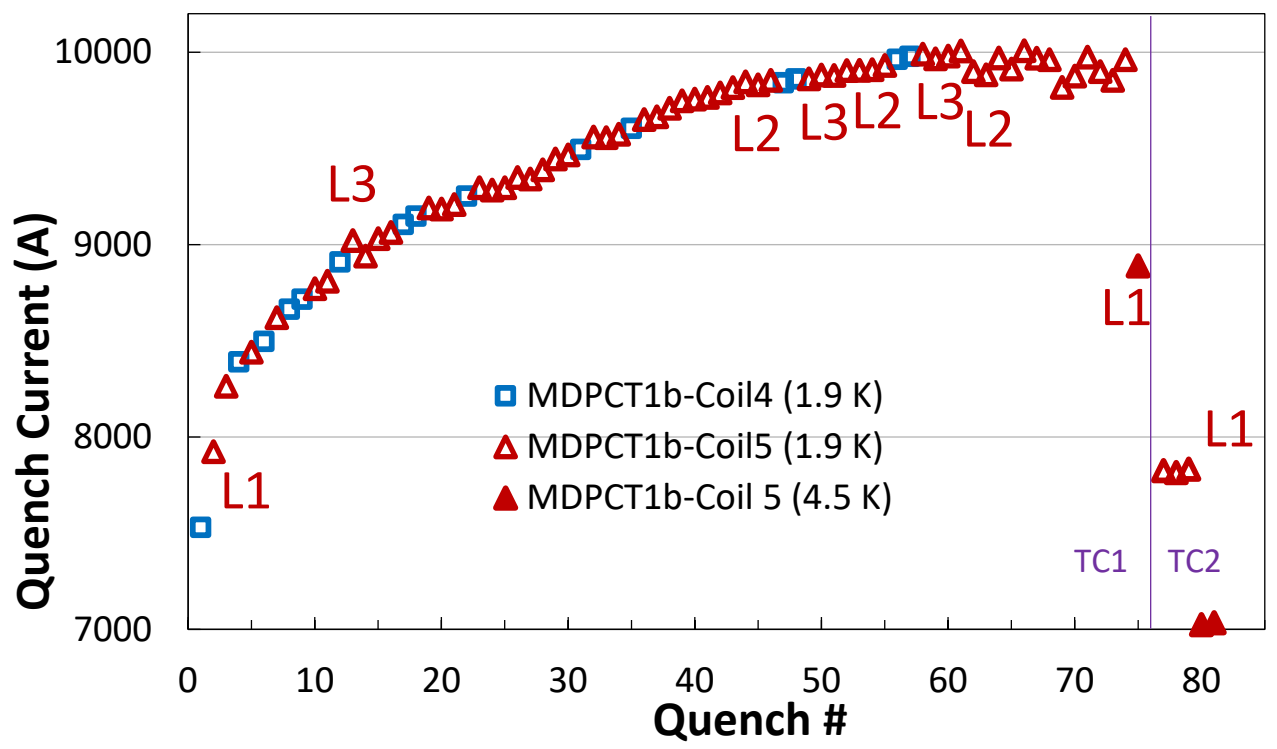

Fig. 3. MDPCT1b training, including different temperatures and thermal cycles. "L*" indicate first occurrences for different quench locations (see later).

## *C. Inner Coils Quenches*

The inner coils, 2 and 3, demonstrated very good performance. The only two quenches in one of the coils were similar and occurred during the initial MDPCT1 training [4], at ~9.7 kA.

In-bore temperature measurements indicated ramp-rate dependent temperature rate increase at 1.9 K. Both "1.9 K" quenches happened closer to 2 K temperature in the conductor. SSL are lower at 2 K than at 1.9 K by about 0.5% – a small correction that could be neglected.

Fig. 4 shows the inner layer quench signatures in the first sensing elements. Both QA gave prompt response to quenches. As seen, the flex-QA also demonstrated excellent signal-to-noise ratio [7]. Having multiple sensors, we were able to determine the quench location and propagation velocity. The location, at the innermost turn, three cm from VT A7 towards the straight section, was consistent between VTs and QA. Quench propagation, based on VT, was 23 m/s.

QA only responded to innermost layer quenches.

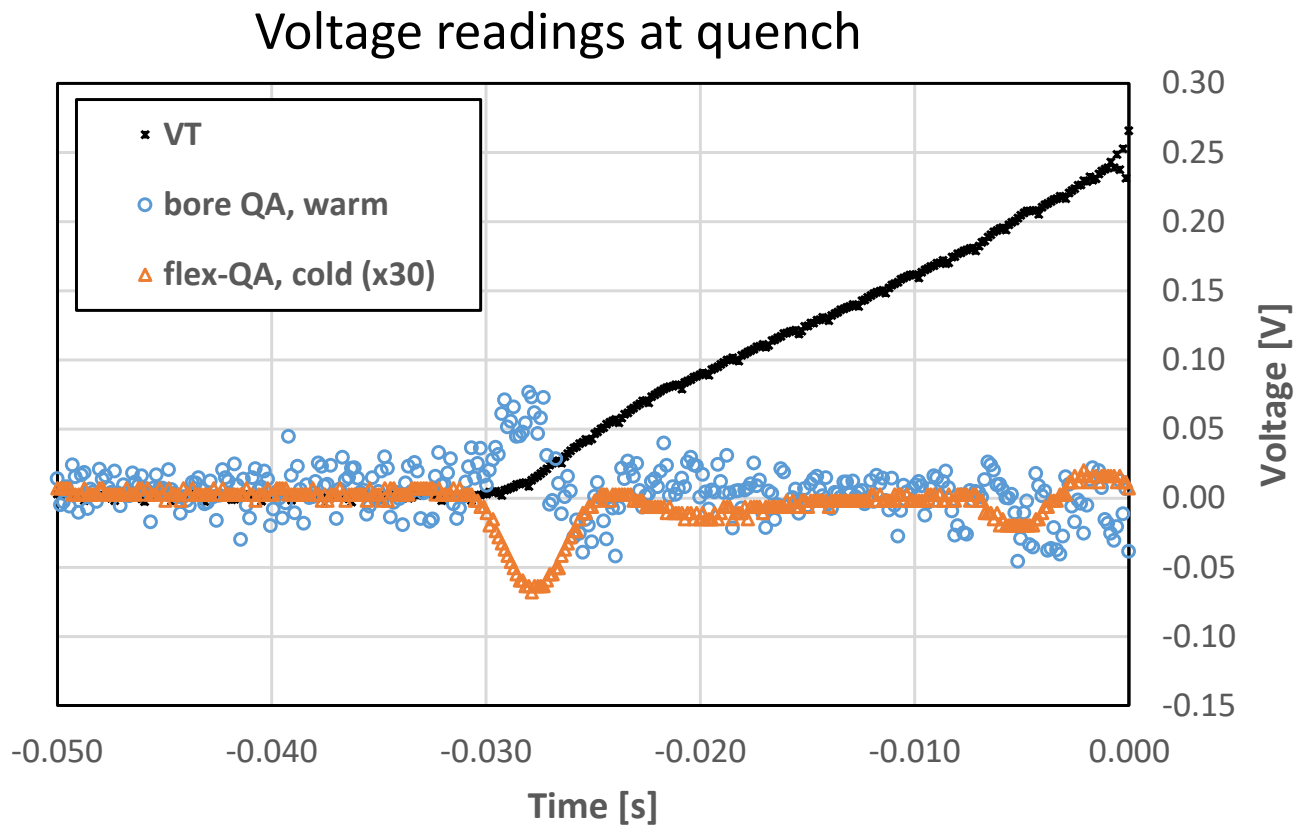

Fig. 4. The second (last) quench in the inner coil: Earliest segment/section of VTs and QAs responding to quench development, quench detection is at 0 s.

## *D. Outer Coils Quenches*

Most of the VT readout traces of the outer coils, 4 and 5, were cut in the initial MDPCT1 testing and we only know that quenches happened in layers 3 and 4 (the same as C and D, Fig. 1), in both coils, in roughly the same proportion. Better location information proved to be unreliable. After repairs, MDPCT1b had fully functional VT taps and we had excellent coverage for quenches. While both coils were training similarly in the first test [5], [10], in MDPCT1b it was coil 5 limiting the performance (Fig. 3). We identified three quench signatures characterizing most of the quenches, Fig. 5.

For all but few quenches we identified quench location to within ~20 cm along the conductor length (some quenches developed in two layers). Coil 5 started quenching in the location L1 at the non-lead end and was followed by different quench-es, with location L3 at the lead-end becoming persistent soon after, Fig. 5 and Fig. 3. Only after 45 quenches the straight section location L2 became active and dominating with increasingly intermittent quenches at L3. Quenches with “erratic” behavior were at location L2. Higher temperature and ramp rate quenches observed after training quenches and all quenches in TC2 happened in the non-lead end location L1. L1 and L2 locations are around the area where the straight and the end sections meet, on one side of the magnet. Most quenches occurred on the same coil side, Fig. 5. About 65% of all MDPCT1b magnet training quenches were in those three locations, and the last quench of different origin and signature in coil 5 was quench 55. During training in TC1, excluding different temperature and ramp rate quenches, there were 28 L3 type quenches, 20 L2 quenches and 3 L1 quenches.

At the end of MDPCT1b testing, we performed dedicated low-noise V-I measurements extracting information about resistive behavior of the weakest VT segment near quench current and behavior of most VT segments at various currents. Results are presented in Fig. 6. We show signals with abnormal behavior and the two most significant ones belong to the segments where spontaneous quenches occur. They develop resistance practically immediately after current start ramping which suggests degradation is likely associated with many strands. As seen on the figure, coil 4 was also showing similar though much less extreme behavior in the same segment(s) with an order of magnitude smaller voltage rate increase vs. current. Some short segments around the affected ones which we were able to measure showed similar trends. The long segments covering outer parts of the layers were too noisy to produce any discernable signal below micro-volt level. No resistance developed in the short inner coil segments.

Investigations of behavior of the weakest VT segment near quench current at 4.5 K demonstrated that it was near critical surface as quench develops. During current increases with as low as 5 A per step and 1 A/s ramp, waiting at plateau for as much as tens of seconds, we monitored voltage behavior. In all cases the voltage trended higher after current increased. The magnet eventually quenched about 100 A above the quench level at 20 A/s ramps (regular ramps). The quench level with the regular ramp rate was reproducible at 1.9 K and 4.5 K within 15 A, for each temperature. No signals were registered by the acoustic sensors within tens of milliseconds before quenches in TC2.

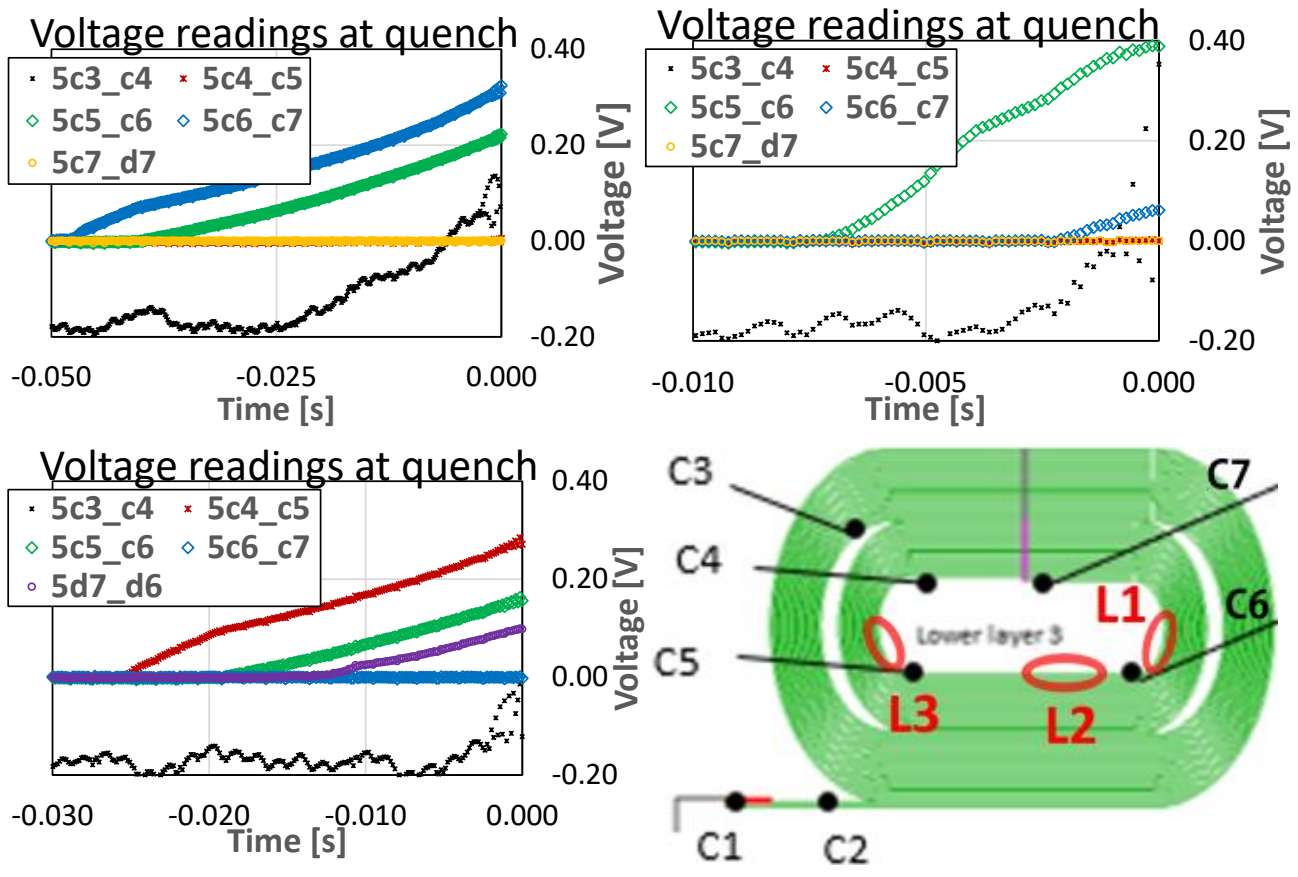

Fig. 5. Voltage segment patterns in MDPCT1b – those three signatures describe most of the quenches. Approximate locations were extracted based on arrival time differences and lack of adjacent activity. From left to right and to bottom: locations/patterns shown are numbered L1, L2 and L3. The figure on lower right shows the quench locations in the coil.

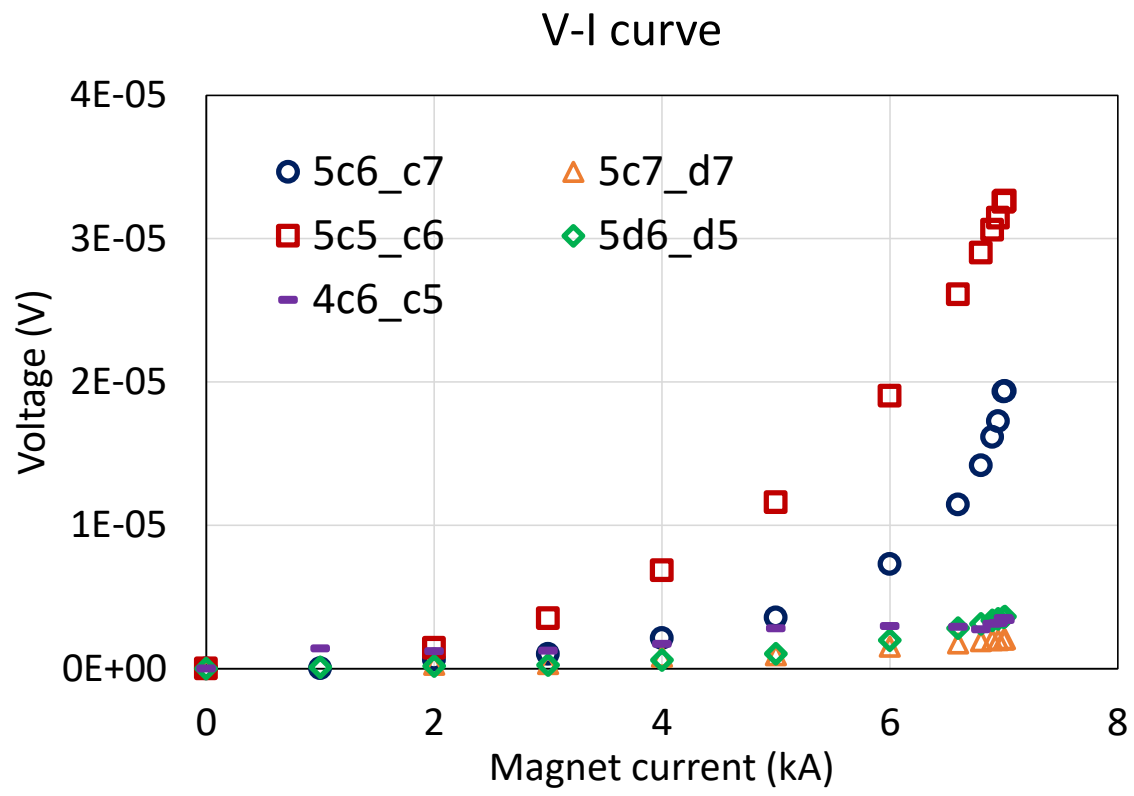

Fig. 6. V-I characteristics of VT segments with developing resistance. At highest currents 5c6_c7 gives 35 μV/A differential increase. The uncertainty on measurement points (RMS over 25 s interval) for the most affected short segments is 30-60 nV.

## III. Magnet Performance

### A. *Conductor Degradation*

Evidence from V-I dependences supports the idea that both outer coils were affected by degradation, at least in the innermost turn, while one of them was damaged much more. Low current resistance rise implies multiple strands were affected but we do not know for certain at which stage. Several segments with this problem suggest a whole area is affected. The fact quenching started at very low fraction of SSL can also hint problems already existed early on. Ultimately the current degradation seen between MDPCT1 and the end of MDPCT1b testing (TC2) was 28%. We speculate damages occurred during the first testing where the end support was later recognized to be sub-optimal [5], [10]. However, an alternative hypothesis about the large and abrupt magnet performance deterioration in TC2 existed, thanks to acoustic data.

### B. *Acoustic Data Analysis*

Most of the quenches in the three types of events discussed in MDPCT1b were linked to little-to-none mechanical activity, as assessed by acoustic signals, milliseconds before or during quenching and nothing changed in that respect after the thermal cycle. Acoustic events in ramps showed activity following the Kaiser effect [11] (magnet becoming “quiet” up to currents already reached), but the thermal cycle largely cleared this memory, i.e. the first ramp in TC2 was “spiky”, Fig. 7 (last ~ 100 s shown), similar to the first ramp in TC1. Acoustic activities of this kind arise mostly from insulation cracking or movements at increasing current/stress levels. Typically, there are hundreds of milliseconds or more of "quietness" between "spikes" when the magnet is subjected to new current levels.

The first ramp of TC2 of MDPCT1b ended with a trip associated to a drop in liquid helium levels at the magnet top due to parallel cryogenic activities. However, a very large mechanical event was detected 8 s before the trip, Fig. 7. This was a unique event.

The time difference Δt between the start of signals from the two acoustic sensors can be determined from Fig. 7 with some uncertainty. The distance L between sensors is known (105 cm) and the velocity (v) of sound large uncertainty is predominantly coming from assumptions about the type of wave propagating (2.5-5 km/s for the solids in the magnet). Then, one can estimate the location of the mechanical event, assuming a point source, as $f_L = [1-\Delta t\ (v/L)]/2$, $f_L$ being the fractional length along the magnet axis where the origin of mechanical disturbance is.

We find the location of the event within about 15 cm along the magnet and it is consistent with the area of the limiting quench locations. However, later examinations of the coil and the magnet did not produce any visual evidence of the large mechanical event observed – no breaking, scratches or other abnormal signs were discovered at any part of the magnet. Thus, we do not have an explanation of what might have caused this disturbance. X-ray/CT-scan photography is the next assessment tool we are planning to use investigating the limiting coil.

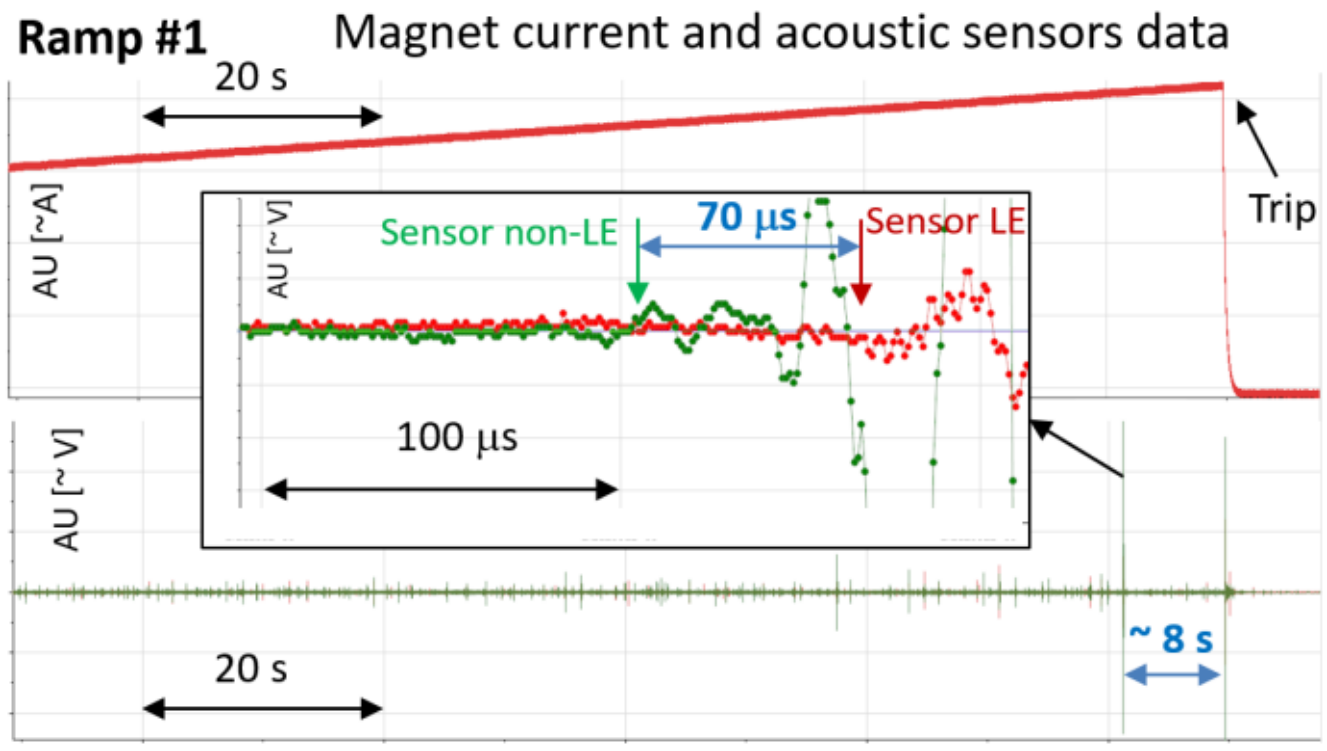

Fig. 7. Current profile for Ramp 1 in TC2 of MDPCT1b (top), acoustic measurements (bottom) at the same timescale and measurements around the beginning of the large spike ~ 8 s before the trip (inset). Y-axes employ arbitrary units (AU), but the large event nearly saturated our DAQ (5 V).

What this large event showed is that longitudinal precision of acoustic event detection can be very good even with many unknowns about parameters of the mechanical wave. It also demonstrated that determination of start of signal can be complicated without additional processing, Fig. 7. It is much more difficult to do the assessment we did with an order of magnitude lower signals, i.e. typical events seen.

### C. *Diagnostics tool assessment*

Diverse data accumulation with sufficient data acquisition rate drastically improves our ability to understand magnet behavior. Following [12], [13, Sec. 9.3.4] we showed how potent low-noise voltage measurement are, although we should have taken data from the beginning of testing and got voltage decay constant measurements. Having high-rate large-area fine-resolution QA measurements would be an excellent addition to instrumentation and the slim flex-QA are the enabling technology. They were also shown to sustain pressures in excess of 200 MPa ([7]), making them suitable for instrumenting in-magnet coil surfaces. Acoustic sensors are a "must" and as a minimum a real-time calibration transducer is necessary to boost their usability. Temperature sensors, possibly in-magnet, do provide excellent insights into behavior when properly positioned and sampled, as demonstrated; an array of them could be a good path forward. All those diagnostic approaches should be part of standard magnet measurements in R&D tests and there are obvious benefits of covering the whole current ramps. High temporal and spatial resolution data of various types covering magnet testing has serious benefits for developments in the field.

We failed to extract data from the outer coil pole strain gauges, and we did not manage yet to get high-rate data from any of the strain gauges. It does have the potential to further shed light on transient behavior, and a run with fiber-optics and temperature sensors in parallel is something of interest before we can pick the optimal approach.

Finally, to realize the potential of all those tools in a magnet test simultaneously is nearly impossible now. The huge obstacle for R&D is the lack of ability to get so many channels at so high data rates out of the pit in a practical way. An elegant solution to this problem is development of "cold" (cryo-) electronics [14], [15] which will have the effect of transforming the way we think about diagnostics in magnet R&D.

## IV. Conclusion

MDPCT1 exhibited quite significant performance decline after reaching record field levels. The outer coils were affected with one of them affected more. This behavior is linked with the lack of sufficient outer coil end support during its initial assembly but is likely also the result of intrinsic deficiencies of this non-stress-managed four-layer magnet structure [10].

Diagnostics helped observing and categorizing multiple events and pinpoint "weak" coil regions. Those are still to be investigated by other means. Sensors of different types were useful to build a more complete picture of the observed behavior of the magnet. In future R&D magnet tests, upgraded and even revolutionized instrumentation is not out of reach and it is needed for optimal use of R&D resources.

## Acknowledgment

The authors would like to thank the APS-TD staff for supporting the magnet testing and instrumentation. Special thanks to our LBNL colleagues for providing feedback, diagnostics hardware, and know-how.